\documentclass[useAMS,usenatbib]{mn2e}

\usepackage{graphicx}
\usepackage{natbib}
\usepackage{subfloat}
\usepackage{amssymb}

\title[Gargantuan chaotic gravitational three-body systems]{Gargantuan chaotic gravitational three-body systems and their irreversibility to the Planck length}
\author[T.~Boekholt, S.~Portegies~Zwart and M.~Valtonen]{T.~C.~N.~Boekholt$^{1,2}$\thanks{E-mail: tjardaboekholt@gmail.com (TB); spz@strw.leidenuniv.nl (SPZ) and mvaltonen2001@yahoo.com (MV)}, S.~F.~Portegies~Zwart$^{3}$\footnotemark[1] and M.~Valtonen$^{4,5}$\footnotemark[1]\\ 
$^{1}$Instituto de Telecomunica\c{c}\~oes, Campus Universit\'ario de Santiago, 3810-193, Aveiro, Portugal \\
$^{2}$Department of Physics, University of Coimbra, 3004-516, Coimbra, Portugal\\
$^{3}$Leiden Observatory, Leiden University, PO Box 9513, 2300 RA, Leiden, The Netherlands\\
$^{4}$Finnish Centre for Astronomy with ESO, University of Turku, FI-20014, Turku, Finland\\
$^{5}$Department of Physics and Astronomy, University of Turku, FI-20014, Turku, Finland
}
 
\begin{document}

\date{Accepted Year Month Day. Received Year Month Day; in original form Year Month Day}

\pagerange{\pageref{firstpage}--\pageref{lastpage}} \pubyear{2015}

\maketitle

\label{firstpage}


\begin{abstract}
Chaos is present in most stellar dynamical systems and manifests itself through the exponential growth of small perturbations. {Exponential divergence drives time irreversibility and increases the entropy in the system. A numerical consequence is that integrations of the N-body problem unavoidably magnify truncation and rounding errors to macroscopic scales.} Hitherto, a quantitative relation between chaos in stellar dynamical systems and the level of irreversibility remained undetermined. In this work we study chaotic three-body systems in free fall initially using the accurate and precise N-body code \texttt{Brutus}, which goes beyond standard double-precision arithmetic. We demonstrate that the fraction of irreversible solutions decreases as a power law with numerical accuracy. This can be derived from the distribution of amplification factors of small initial perturbations. Applying this result to {systems consisting of three massive black holes} with zero total angular momentum, we conclude that up to five percent of such triples would require an accuracy of smaller than the Planck length in order to produce a time-reversible solution, thus rendering them fundamentally unpredictable.
\end{abstract}




\section{Introduction}


Chaos is an inherent property of most dynamical systems in the universe, ranging from small bodies in the solar system \citep[e.g.][]{1984Icar...58..137W,2016MNRAS.461.3576B,2018Icar..305..250C}, {small stellar systems} \citep[e.g.][]{1983ApJ...268..319H,2014ApJ...785L...3P,2015ComAC...2....2B, 2018CNSNS..61..160P, 2018MNRAS.476..336L, 2019arXiv190905272S}, star clusters \citep[e.g.][]{1964ApJ...140..250M,1993ApJ...415..715G} and galaxies \citep[e.g.][]{2000chun.proc..229V}. 
The main signature of chaos is the exponential sensitivity to small changes in the initial conditions, which is quantified by the e-folding time scale {within some finite time interval, i.e. the finite time Lyapunov time scale \citep{1991pscn.proc...47H}.} 

The exponential sensitivity in N-body systems has both physical and numerical consequences. From a physical point of view, the rate of growth of perturbations determines the stability of a system. {Such studies are well-known for the solar system, whose Lyapunov time scale is about 5 Myr \citep{1989Natur.338..237L}. Due to observational uncertainties in the orbital elements of the planets, we can only predict the future evolution of the solar system for a few million years, warranting a statistical study of its stability over Gyr time scales \citep{1989Natur.338..237L, 1992Sci...257...56S, 2002MNRAS.336..483I, 2007NatPh...3..689H}. Hence, in contrast to regular and stable systems, high precision in the initial conditions is crucial for accurate modelling of chaotic systems. } 
  
{In few-body stellar dynamical systems, it was first shown by \citet{1964ApJ...140..250M} that two nearby trajectories in phase space tend to diverge exponentially. ``The divergence of the two trajectories from each other is a manifestation of the macroscopic irreversibility'' and ``the rate of divergence yields information on the rate of entropy production'' \citep{1964ApJ...140..250M}. This rate is linear with time, because the rate of divergence is exponential, and the entropy proportional to the logarithm of the increasing phase space volume. The presence of chaos and macroscopic irreversibility can be related to the arrow of time, in the sense that it points in the direction of increasing entropy. Thus the arrow of time points in the direction of diverging trajectories rather than converging ones. This leads to the idea that in a world consisting of only three bodies, there would already be a definite direction for the arrow of time \citep{2008MNRAS.388..965L}.}

{From a numerical point of view,} errors in N-body simulations also act as small perturbations to the system, and their subsequent exponential magnification causes the solution to eventually  diverge onto a completely different trajectory after only a few Lyapunov time scales. {The calculated system is not causally related to its initial condition anymore, in the same way a physical system is \citep{1964ApJ...140..250M}.} 
{This raises suspicions on the reliability of N-body simulations. The common assumption is that approximate results from N-body simulations are valid in a statistical sense \citep{1993ApJ...415..715G}. Empirically this has been shown to be the case for certain specific N-body sytems \citep[e.g.][]{2014ApJ...785L...3P, 2015ComAC...2....2B, 2018CNSNS..61..160P}, but a sound theoretical basis is still missing. Our trust in N-body simulations can potentially be made more robust if it can be shown that the ``numerically diverged trajectory'' still has some physical connection to the initial condition space under consideration, but to a slightly different initial realization than the one used to start the simulation. In other words, we are still calculating physical trajectories, but to randomized initial conditions \citep{1986LNP...267..212D}. This process can be demonstrated if it can be shown that approximate solutions have ``shadow orbits'' \citep{1992MNRAS.259..505Q, 2010MNRAS.407..804U}. Such orbits remain close to the approximate trajectory for a time much longer than the Lyapunov time, but which have a physical connection to the initial condition space of the N-body problem under consideration.}  

{Alternatively, one can apply brute-force computing power to try and reduce the magnitude of numerical errors. A robust way to test the accuracy of a specific N-body simulation is by performing a reversibility test. }
Since Newton's equations of motion are time reversible, a forward integration followed by a backward integration of the same time should recover the initial realization of the system (albeit with a sign difference in the velocities). The outcome of a reversibility test is thus exactly known. In practice, reversibility in simulations of chaotic systems is very difficult to achieve due to 1) {exponential growth of perturbations due to chaos}, and 2) irreversible numerical errors. Time reversibility can be forced by using integer arithmetic, but this does not garantuee the solution is also accurate.   


Recently, \citet{2018CNSNS..61..160P} obtained a reversible solution to the Pythagorean problem \citep{1913AN....195..113B, 1967AJ.....72..876S, 1994CeMDA..58....1A}. {This is a classic example of a three-body system in free fall initially exhibiting a prolonged chaotic triple interaction, and an eventual break up into a permanent and unbound binary-single pair.} They applied the \texttt{Brutus} N-body code and the method of convergence in which the accuracy and precision of the integration is systematically increased until convergence of the solution to the first few decimal places \citep{2015ComAC...2....2B}.
Although \texttt{Brutus} is not formally time reversible, they manage to retrieve the initial condition to the first 10 decimal places in each coordinate of each body in the final snapshot. Whereas the forward integration is subject to exponential divergence, the backward integration is subject to exponential convergence to the initial size of the perturbation over nine orders of magnitude. This behavior was called {\it definitive reversibility} by \citet{2018CNSNS..61..160P}. 


We extend the initial condition space from the Pythagorean problem to the homology map of \citet{1967AZh....44.1261A, 1968SvA....11.1006A} {(see also \citet{1991A&A...252..410A,1991CeMDA..51....1A,1994CeMDA..60..365A,1995CeMDA..62..335T,2014ARep...58..756M, 2016ARep...60.1083O})}. {For a definition and visualization of this map, see Fig.~1 of \citet{2008MNRAS.388..965L}}. {The Agekyan-Anosova map consists of every equal-mass triple system configuration with zero initial velocities (after potentially rescaling or rotating the system).} The initial conditions thus specify three-body systems in free fall initially with varying initial mutual separations. Such trajectories {may} closely approach a triple collision, which are notoriously challenging to solve, even using regularization techniques. {After such close triple approaches, the triple can break up, or alternatively continue its evolution in a prolonged, chaotic and resonant \citep{1983ApJ...268..319H} interaction. }
Reversibility tests for the ``homology map'' have been performed by \citet{2008MNRAS.388..965L}. They find that about half of the systems are reversible and the other half remains irreversible, regardless how much computer time you spend on the problem. They conclude that half of the three-body systems are so chaotic that they cannot be solved numerically \citep{2006Valtonen}. This is also corroborated by results from \citet{1986LNP...267..212D} who measure amplification factors of small initial perturbations of up to $10^{150}$. 
However, the fraction of irreversible solutions can potentially be reduced if one goes beyond standard double-precision, i.e. using arbitrary-precision arithmetic. 

\section{Results}

\begin{figure*}
\centering
\begin{tabular}{cc}
\includegraphics[height=0.5625
\textwidth,width=0.45\textwidth]{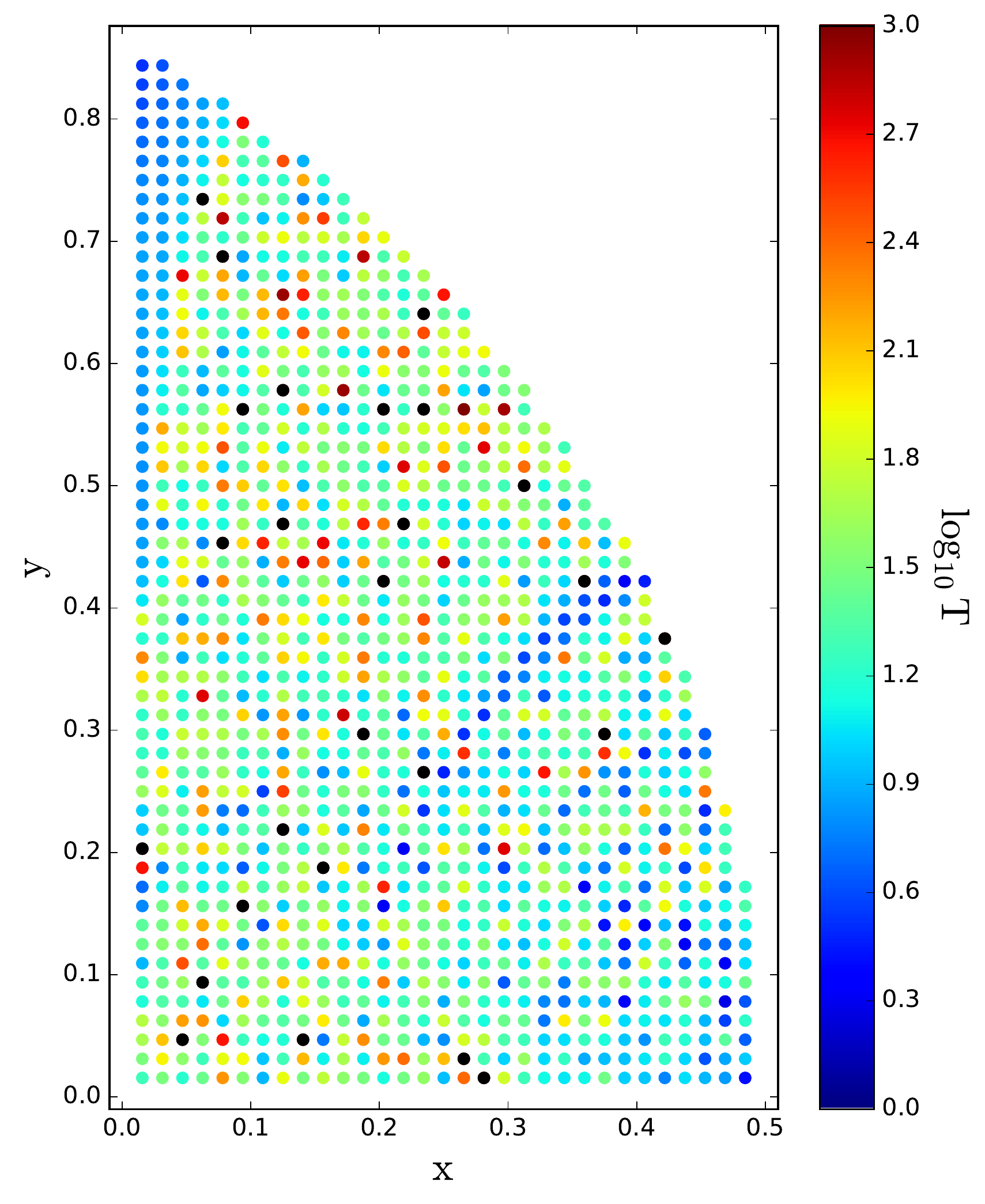} &
\includegraphics[height=0.5625
\textwidth,width=0.45\textwidth]{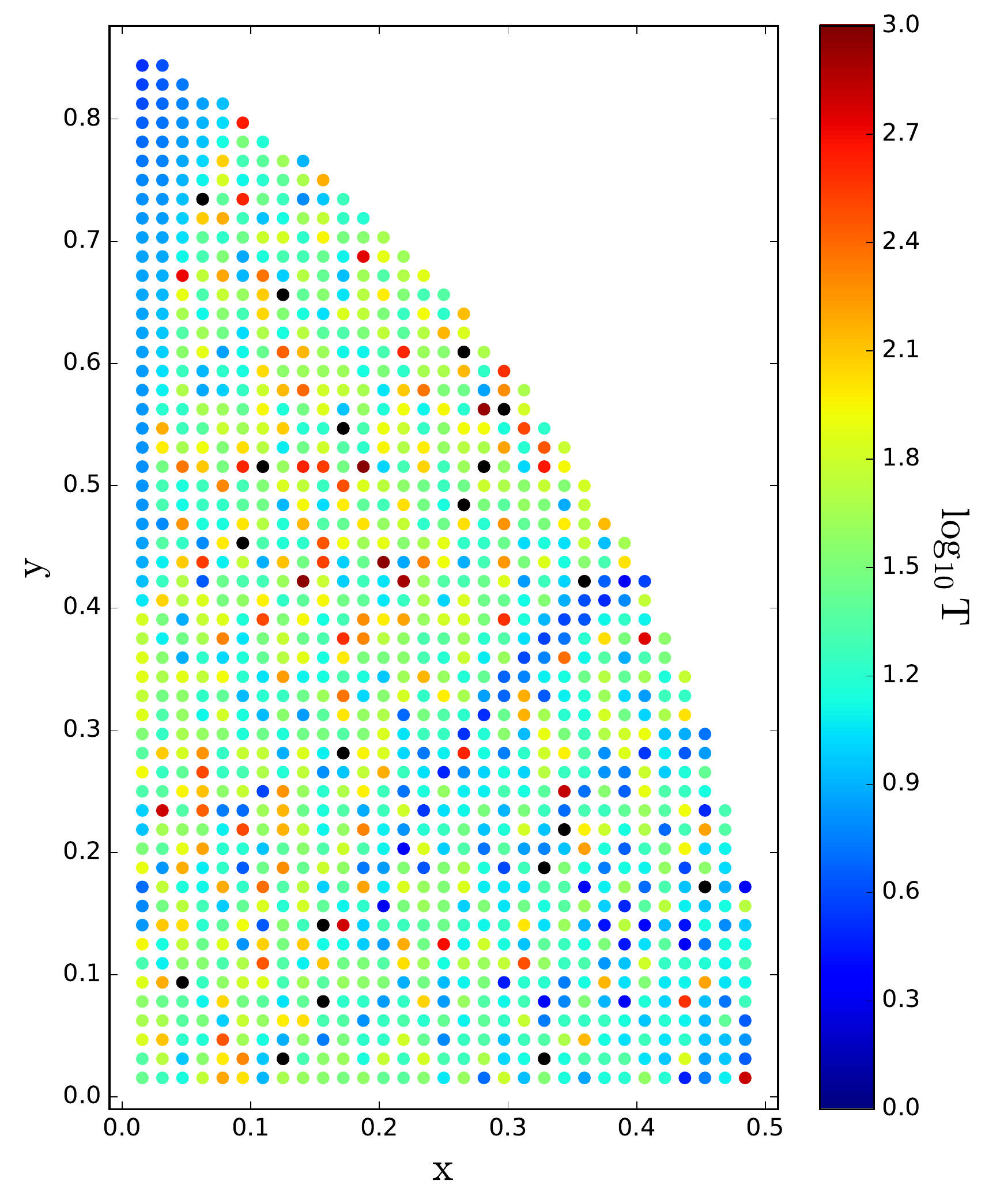} \\
\end{tabular}  
\caption{ Duration of the triple interaction as a function of initial condition in the Agekyan-Anosova map \citep[a higher resolution map can be found in Fig. 1 of ][]{2008MNRAS.388..965L}. We show the ``lifetime map'' for a numerical accuracy of $\epsilon = 10^{-6}$ (left) and $\epsilon = 10^{-70}$ (right). Black dots represent systems that live for longer than a 1000 time units. {Even though the same initial condition in the two maps can give very different lifetimes, the two maps are statistically indistinguishable according to a Kolmogorov-Smirnoff test.} }
\label{fig:maps}
\end{figure*}

We perform reversibility tests for the Agekyan-Anosova map \citep{1967AZh....44.1261A, 1968SvA....11.1006A, 2008MNRAS.388..965L} using the arbitrary-precision N-body code \texttt{Brutus} \citep{2015ComAC...2....2B}. We control the accuracy by varying the Bulirsch-Stoer tolerance \citep{springerlink:10.1007/BF01386092}, $\epsilon$, and fix the arbitrary-precision word-length to 1024 bits (about 300 decimal places). More detail on the methods is given in Appendix~A. 

{The main idea of our experiment is the following. Each triple system has a certain escape time, which is the time it takes for the triple to break up into a permanent and unbound binary-single configuration. Given a numerical accuracy, $\epsilon$, there is also a tracking time, which is the time that the numerical solution is still close to the physical trajectory that is connected to the initial condition. If the tracking time is shorter than the escape time, then the numerical solution has diverged from the physical solution, and as a consequence, it has become time irreversible. Only the systems with the smallest amplifications factors will pass the reversibility test. However, by systematically increasing the numerical accuracy (decreasing $\epsilon$), we aim to increase the tracking time of each system. An increasing fraction of systems will obtain a tracking time exceeding its escape time, thus gradually decreasing the fraction of irreversible solutions. } 

In Fig.~1 we present our low resolution Agekyan-Anosova map, where we plot the lifetime of the triple system as a function of initial condition. {The triple lifetime is defined as the duration of the triple interaction until a permanent and unbound binary-single configuration is reached}. When comparing the least accurate ($\epsilon = 10^{-6}$) and the most accurate ($\epsilon = 10^{-70}$) maps, we observe that there are ``microscopic'' differences. However, in a ``macroscopic'' sense,  the maps look similar. This is confirmed by performing a Kolmogorov-Smirnoff test, which gives a p-value of 0.72. {This implies that we cannot reject the hypothesis that the two distributions are statistically indistinguishable.} Therefore, it seems that for the Agekyan-Anosova map, approximate computations are nevertheless reliable in a statistical sense \citep{1993ApJ...415..715G}. {This is another example of the concept of ``{\it nagh-Hoch}'' \citep{2018CNSNS..61..160P}, which refers to the ``similar appearance'' of statistical distributions, which are obtained with different numerical precisions.\footnote{The term ``Nagh Hoch'' was first defined by \citet{2018CNSNS..61..160P} and comes from the Klingon dictionary meaning ``similar appearance'' or ``set in stone''.}}

In Fig.~2 we plot the fraction of irreversible solutions as a function of numerical accuracy, i.e. the Bulirsch-Stoer tolerance, $\epsilon$. For an accuracy close to double-precision, the fraction of irreversible solutions is about half, consistent with the results of \citet{2008MNRAS.388..965L}. By increasing the numerical accuracy beyond machine-precision, we demonstrate that we are able to further decrease the fraction of irreversible solutions. The data is accurately fitted by a power law, given by

\begin{equation}
\log_{10} f_{\rm{irr}} = \alpha \log_{10}\,\epsilon + \beta,
\label{eq:2}
\end{equation}  

\noindent with $\alpha=0.029 \pm 0.001$ and $\beta=0.15 \pm 0.04$. 

In Fig.~3 we plot the distribution function of the amplification factors of  small initial perturbations. This quantity is defined as the {Euclidean norm of the distance in position space} between the forward and backward integration as a function of time\footnote{This is similar to the phase space distance \citep[Eq.2]{1964ApJ...140..250M}, but only considering the position coordinates.  }. In a perfectly time reversible integration the {norm} should remain zero, but during a numerical integration it will grow exponentially. The final distance divided by the initial distance (after a single integration step) is the amplification factor, $A$. We find that the distribution is accurately fitted by a power law, given by 

\begin{equation}
\log_{10}\frac{df}{d\log_{10} A} = \gamma \log_{10} A + \delta,
\label{eq:1}
\end{equation}

\noindent with $\gamma=-0.0270 \pm 0.0008$ and $\delta=-1.20 \pm 0.01$. If this relation could be extrapolated, this would imply that for a very high sampling of the Agekyan-Anosova map, there should be some systems with amplification factors exceeding $10^{100}$, which would take a long time to calculate up to convergence. The average wall-clock time of our simulations was about 7 hours, with the longest run taking 1 month to complete.

{We observe a very large difference in the ranges of the axes in Fig. 2 and 3. Whereas the fractions vary over two orders of magnitude, the accuracy and amplification factors vary over 70 orders of magnitude in Figs. 2 and 3 respectively. The slope is very small and only resolvable due to the very high accuracy and precision of the \texttt{Brutus} code. The results are inconsistent with a flat curve, which intuitively makes sense since with higher numerical accuracy we expect to reduce the fraction of irreversible solutions.}

\begin{figure}
\centering
\begin{tabular}{c}
\includegraphics[height=0.36\textwidth,width=0.45\textwidth]{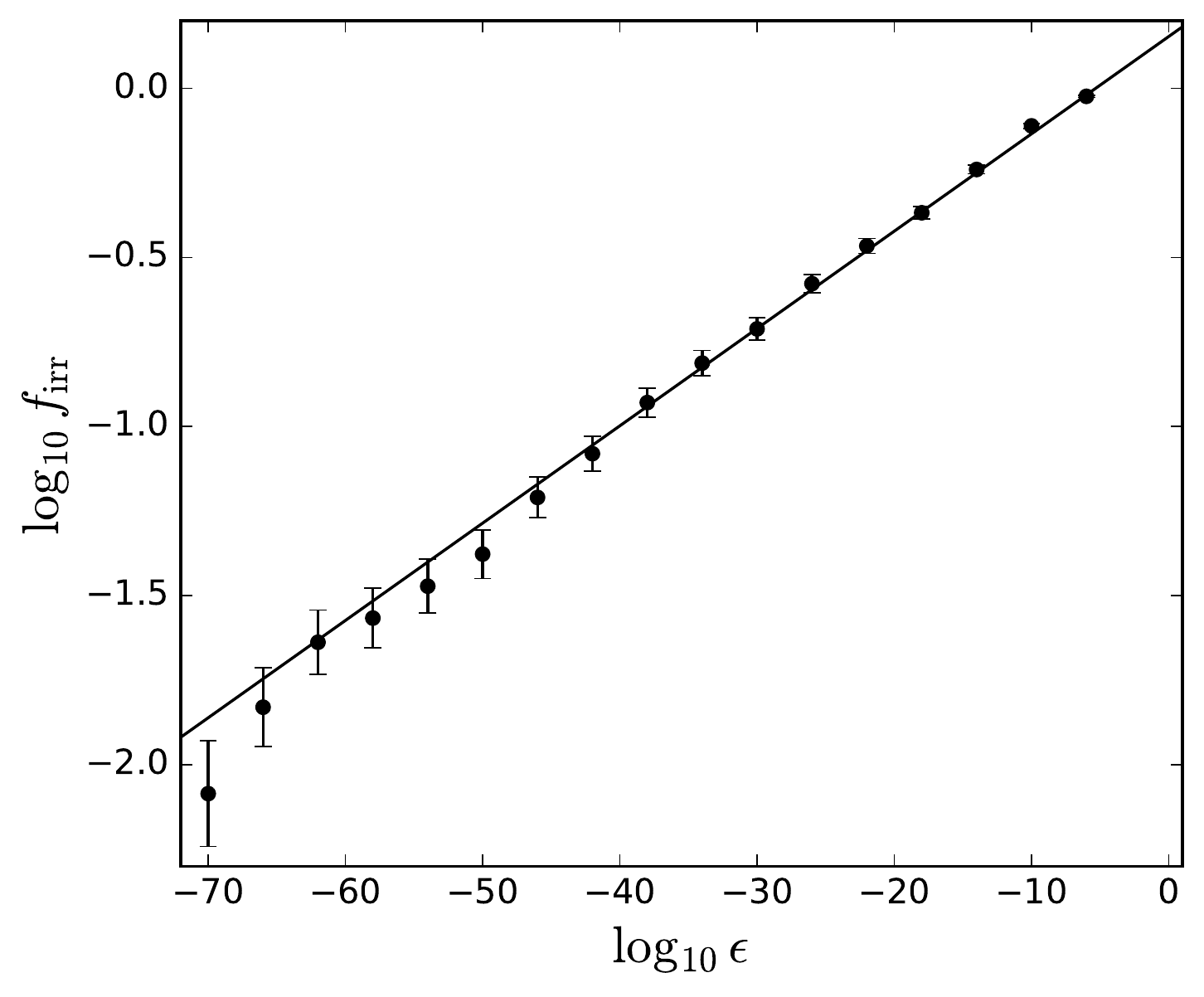} \\
\end{tabular}  
\caption{ Fraction of irreversible solutions, $f_{\rm{irr}}$, as a function of numerical accuracy, $\epsilon$. The power law fit gives: $\log_{10}\,f_{\rm{irr}} = \alpha \log_{10}\,\epsilon + \beta$, with $\alpha=0.029 \pm 0.001$ and $\beta=0.15 \pm 0.04$.  }
\label{fig:fraction_vs_epsilon}
\end{figure}

\begin{figure}
\centering
\begin{tabular}{c}
\includegraphics[height=0.36\textwidth,width=0.45\textwidth]{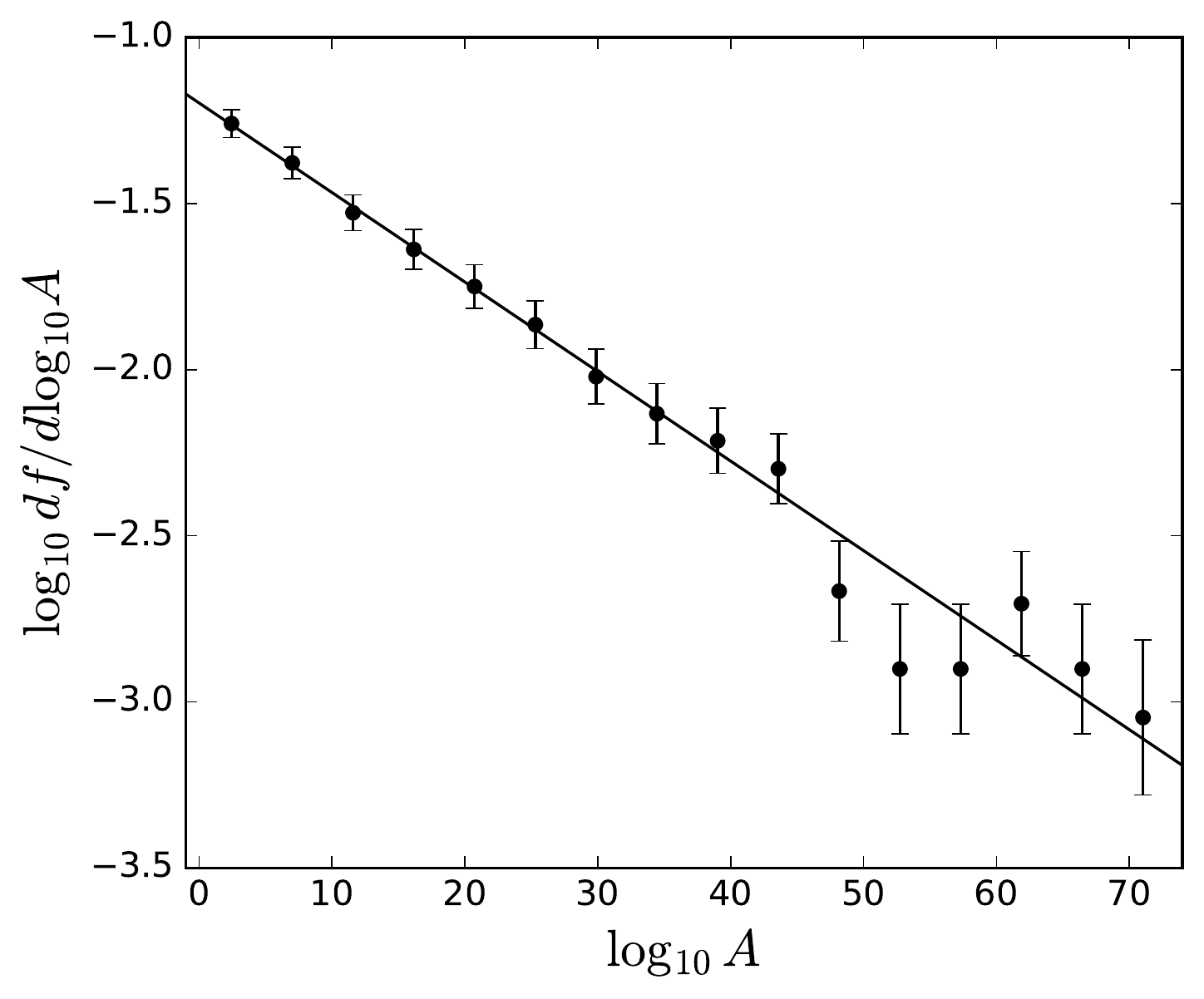} \\
\end{tabular}  
\caption{ Distribution of amplification factors. The jackknife estimated errorbars increase towards large values of $A$ due to the decrease in number of systems. The power law fit gives: $\log_{10}\,df/d\log_{10}A = \gamma \log_{10}A + \delta$, with $\gamma=-0.0270 \pm 0.0008$ and $\delta=-1.20 \pm 0.01$.  }
\label{fig:fraction_vs_A}
\end{figure}

Finally, we show in Fig.~4 that there is only a mild dependence of the required numerical accuracy, $\epsilon$, on the closest encounter between any two bodies during the simulation.  
This is because a close two-body encounter with a small perturbation from the third body is well approximated by a Keplerian orbit. It is the close encounter plus a large third body perturbation that may lead to loss of numerical accuracy.
The amplification factor is determined by both the lifetime of the triple system, and the magnitude of the finite-time Lyapunov exponent. It remains an open question what determines the rate of exponential growth and transitions in the growth (see for example Fig.~1 of \citet{2018CNSNS..61..160P}). The exponential sensitivity, or ``the butterfly effect'', can be explained in terms of separatrix crossings \citep{2008LNP...760...59M}. A trajectory in phase space approaches a separatrix, which divides regions of libration and circulation. It might succeed in crossing the separatrix to a new region in phase space, with potentially different chaotic properties. However, a nearest-neighbour trajectory might fail to cross the separatrix and remain in its current region of phase space, resulting in an exponential magnification of its separation in phase space from the other trajectory. A detailed study of the relation between orbital geometries and the rate of exponential growth of small perturbations will be presented elsewhere.


\begin{figure}
\centering
\begin{tabular}{c}
\includegraphics[height=0.36\textwidth,width=0.45\textwidth]{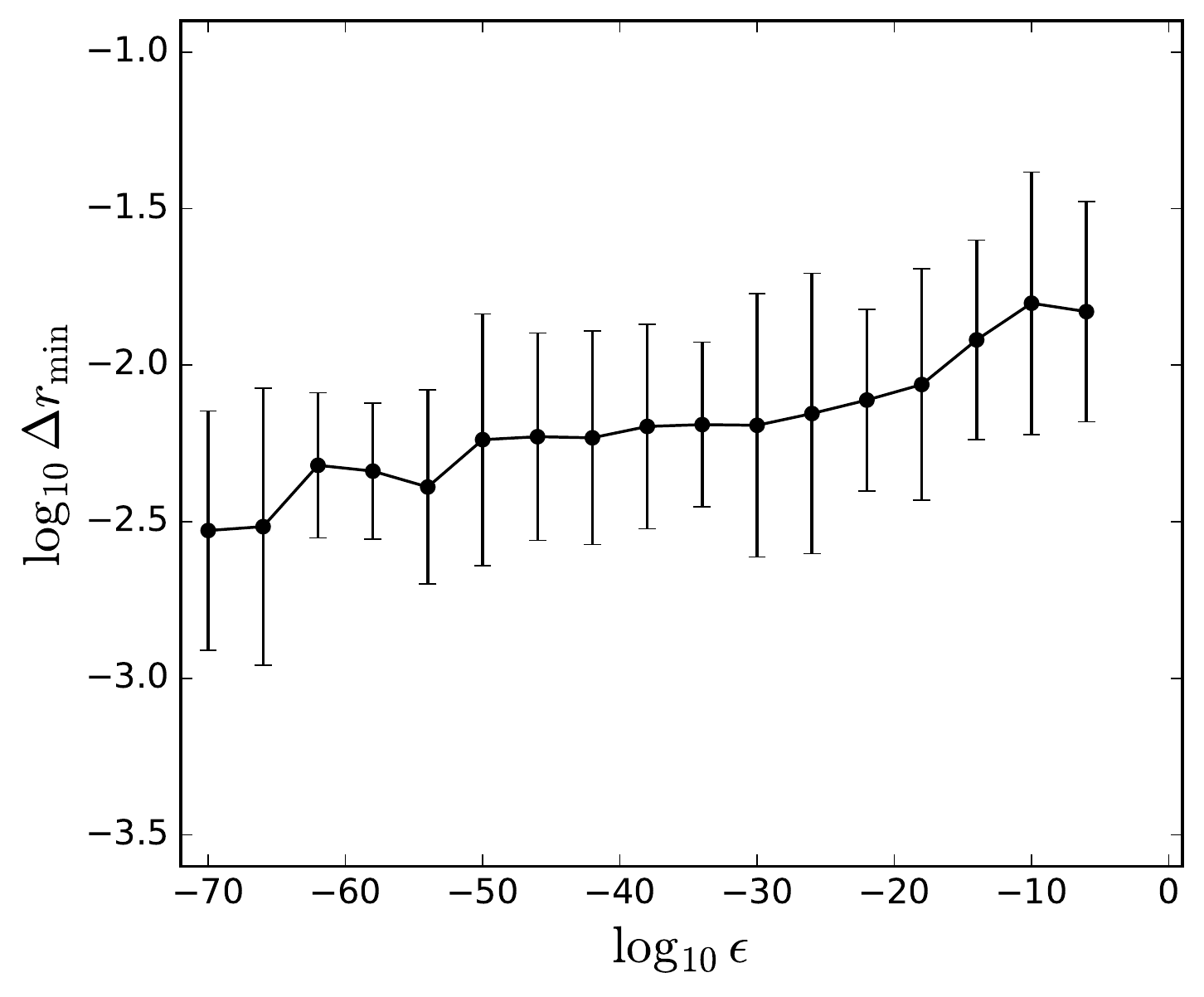} \\
\end{tabular}  
\caption{ Minimum separation, $\Delta r_{\rm{min}}$, between two bodies during a simulation, which required a Bulirsch-Stoer tolerance, $\epsilon$, in order to pass the reversibility test. The data points and errorbars correspond to the average and standard deviation. }
\label{fig:drminmin_vs_epsilon}
\end{figure}

{The errorbars in Figs. 2 and 3 result from a finite sampling of the Agekyan-Anosova map. The exact fraction of irreversible solutions as a function of accuracy is obtained when sampling the map with infinite resolution. However, this is impractical, and instead we generate a sample of {1212} initial conditions by overlaying a uniform grid on the map with a grid-spacing of 0.015625. The uncertainty due to the finite sample is estimated by the jackknife resampling method. The errorbars in Fig. 4 are much larger, simply because the variation in the minimal distance between two bodies varies a lot among different simulations with the same accuracy. This shows that the exponential growth is not driven by close two-body encounters, but rather by a prolonged phase of strong three-body encounters.  }

\section{Discussion}

The absolute values of the coefficients $\alpha$ and $\gamma$ are statistically indistinguishable. This suggests that a relation exists between the distribution of amplification factors and the fraction of irreversible solutions. Given a certain value of the Bulirsch-Stoer tolerance, $\epsilon$, we can only resolve amplification factors of order $\epsilon^{-1}$, i.e. with $\epsilon = 10^{-10}$ amplification factors of order $A=10^{10}$ can be resolved. Hence, the fraction of irreversible solutions equals the fraction of systems with an $A > \epsilon^{-1}$, i.e. $f_{\rm{irr}}\left( \epsilon \right) = F\left( A > \epsilon^{-1}\right)$. {Thus, the fraction of irreversible solutions for a given numerical accuracy, $\epsilon$, is determined by the distribution of amplification factors of the systems in the initial condition map. In Fig.~3 we show that the distribution of amplification factors is also a power law. In Appendix~B we demonstrate that the two power laws in Figs. 2 and 3 are related. }

{In this study we limit our initial conditions to three-body systems in free fall, i.e. with zero angular momentum. The power law indices measured in Sec. 2 reflect these initial conditions and the way the homology map was sampled. It is likely that for a different set of initial conditions, such as a Plummer distribution with non-zero angular momentum, the power law indices might be different. Nevertheless, large amplification factors can still be expected for some non-zero angular momentum systems, since the amplification factor is determined not only by the finite-time Lyapunov exponent (through close triple encounters), but also by the duration of the triple interaction, which is considerably longer for systems with a larger angular momentum \citep[e.g.][Fig. 7]{2015ComAC...2....2B}. An example for a binary-single scattering experiment is given in Fig.~4 of \citet{1986LNP...267..212D} who measure an amplification factor of about $10^{40}$ over a time interval of about 5400 N-body time units. } 

In the limit of infinite accuracy ($\epsilon \rightarrow 0$) we retrieve the microscopic time-reversibility of Newton's equations of motion. In the presence of perturbations of size $\epsilon$, whether numerical or physical, a fraction of systems becomes irreversible. As a concrete application of our result, we consider three black holes, each of a million solar masses, and initially separated from each other by roughly one parsec. Such a configuration is not uncommon among supermassive black holes in the concordance model of cosmology and hierarchical galaxy formation \citep{2010MNRAS.402.2308A}. Here we will focus specifically on the subset of triples which approach zero angular momentum, consistent with the systems studied in this work.
From Fig.~4 we estimate that the closest approach between any two black holes is on average between $10^{-2.5}$ and $10^{-2}$ parsec, during which the Newtonian approximation still holds. A parsec equals $10^{51}$ Planck lengths. Hence, from Eq.~1 and Fig.~2, we estimate that {up to} 5 percent of triples with zero angular momentum are irreversible up to the Planck length, thus rendering them fundamentally unpredictable. 


\section{Acknowledgements}

We thank Rosemary Mardling for in-depth discussions about the results in this Letter and providing us with feedback for improving the  presentation. We also thank Nathan Leigh for carefully reading our manuscript and providing us with constructive feedback. We thank Sverre Aarseth and Lee Smolin for interesting discussions on chaos and high-precision simulations.  TB acknowledges support from ENGAGE SKA RI, grant POCI-01-0145-FEDER-022217, funded by COMPETE 2020 and FCT, Portugal. {This work is funded by FCT/MEC through national funds and when applicable co-funded by FEDER – PT2020 partnership agreement under the project UID/EEA/50008/2019.} The calculations ware performed using the LGMII (NWO grant \#621.016.701).


\bibliographystyle{mn2e} 
\bibliography{agekyan.bib}      

\begin{thebibliography}{}

\bibitem[\protect\citeauthoryear{{Aarseth}, {Anosova}, {Orlov} \&
  {Szebehely}}{{Aarseth} et~al.}{1994}]{1994CeMDA..58....1A}
{Aarseth} S.~J.,  {Anosova} J.~P.,  {Orlov} V.~V.,    {Szebehely} V.~G.,  1994,
  Celestial Mechanics and Dynamical Astronomy, 58, 1

\bibitem[\protect\citeauthoryear{{Agekyan} \& {Anosova}}{{Agekyan} \&
  {Anosova}}{1967}]{1967AZh....44.1261A}
{Agekyan} T.~A.,  {Anosova} Z.~P.,  1967, Astronomicheskii Zhurnal, 44, 1261

\bibitem[\protect\citeauthoryear{{Agekyan} \& {Anosova}}{{Agekyan} \&
  {Anosova}}{1968}]{1968SvA....11.1006A}
{Agekyan} T.~A.,  {Anosova} Z.~P.,  1968, Astronomicheskii Zhurnal, 11, 1006

\bibitem[\protect\citeauthoryear{{Amaro-Seoane}, {Sesana}, {Hoffman},
  {Benacquista}, {Eichhorn}, {Makino} \& {Spurzem}}{{Amaro-Seoane}
  et~al.}{2010}]{2010MNRAS.402.2308A}
{Amaro-Seoane} P.,  {Sesana} A.,  {Hoffman} L.,  {Benacquista} M.,  {Eichhorn}
  C.,  {Makino} J.,    {Spurzem} R.,  2010, Monthly Notices of the Royal
  Astronomical Society, 402, 2308

\bibitem[\protect\citeauthoryear{{Anosova}, {Orlov} \& {Aarseth}}{{Anosova}
  et~al.}{1994}]{1994CeMDA..60..365A}
{Anosova} J.~P.,  {Orlov} V.~V.,    {Aarseth} S.~J.,  1994, Celestial Mechanics
  and Dynamical Astronomy, 60, 365

\bibitem[\protect\citeauthoryear{{Anosova}}{{Anosova}}{1991}]{1991CeMDA..51....1A}
{Anosova} Z.~P.,  1991, Celestial Mechanics and Dynamical Astronomy, 51, 1

\bibitem[\protect\citeauthoryear{{Anosova} \& {Nebukin}}{{Anosova} \&
  {Nebukin}}{1991}]{1991A&A...252..410A}
{Anosova} Z.~P.,  {Nebukin} A.~V.,  1991, Astronomy and Astrophysics, 252, 410

\bibitem[\protect\citeauthoryear{{Boekholt} \& {Portegies Zwart}}{{Boekholt} \&
  {Portegies Zwart}}{2015}]{2015ComAC...2....2B}
{Boekholt} T.,  {Portegies Zwart} S.,  2015, Computational Astrophysics and
  Cosmology, 2, 2

\bibitem[\protect\citeauthoryear{{Boekholt}, {Pelupessy}, {Heggie} \&
  {Portegies Zwart}}{{Boekholt} et~al.}{2016}]{2016MNRAS.461.3576B}
{Boekholt} T.~C.~N.,  {Pelupessy} F.~I.,  {Heggie} D.~C.,    {Portegies Zwart}
  S.~F.,  2016, Monthly Notices of the Royal Astronomical Society, 461, 3576

\bibitem[\protect\citeauthoryear{{Bulirsch} \& {Stoer}}{{Bulirsch} \&
  {Stoer}}{1964}]{springerlink:10.1007/BF01386092}
{Bulirsch} R.,  {Stoer} J.,  1964, Numerische Mathematik, pp 413--427

\bibitem[\protect\citeauthoryear{{Burrau}}{{Burrau}}{1913}]{1913AN....195..113B}
{Burrau} C.,  1913, Astronomische Nachrichten, 195, 113

\bibitem[\protect\citeauthoryear{{Correia}}{{Correia}}{2018}]{2018Icar..305..250C}
{Correia} A.~C.~M.,  2018, Icarus, 305, 250

\bibitem[\protect\citeauthoryear{{Dejonghe} \& {Hut}}{{Dejonghe} \&
  {Hut}}{1986}]{1986LNP...267..212D}
{Dejonghe} H.,  {Hut} P.,  1986, in {Hut} P.,  {McMillan} S.~L.~W.,  eds, The
  Use of Supercomputers in Stellar Dynamics Vol.~267 of Lecture Notes in
  Physics, Berlin Springer Verlag, {Round-Off Sensitivity in the N-Body
  Problem}.
p.~212

\bibitem[\protect\citeauthoryear{{Goodman}, {Heggie} \& {Hut}}{{Goodman}
  et~al.}{1993}]{1993ApJ...415..715G}
{Goodman} J.,  {Heggie} D.~C.,    {Hut} P.,  1993, Astrophysical Journal, 415,
  715

\bibitem[\protect\citeauthoryear{{Hayes}}{{Hayes}}{2007}]{2007NatPh...3..689H}
{Hayes} W.~B.,  2007, Nature Physics, 3, 689

\bibitem[\protect\citeauthoryear{{Heggie}}{{Heggie}}{1991}]{1991pscn.proc...47H}
{Heggie} D.~C.,  1991, in {Roeser} S.,  {Bastian} U.,  eds, Predictability,
  Stability, and Chaos in N-Body Dynamical Systems {Chaos in the N-body problem
  of stellar dynamics.}.
pp 47--62

\bibitem[\protect\citeauthoryear{{Heggie} \& {Mathieu}}{{Heggie} \&
  {Mathieu}}{1986}]{1986LNP...267..233H}
{Heggie} D.~C.,  {Mathieu} R.~D.,  1986, in {Hut} P.,  {McMillan} S.~L.~W.,
  eds, The Use of Supercomputers in Stellar Dynamics Vol.~267 of Lecture Notes
  in Physics, Berlin Springer Verlag, {Standardised Units and Time Scales}.
p.~233

\bibitem[\protect\citeauthoryear{{Hut} \& {Bahcall}}{{Hut} \&
  {Bahcall}}{1983}]{1983ApJ...268..319H}
{Hut} P.,  {Bahcall} J.~N.,  1983, Astrophysical Journal, 268, 319

\bibitem[\protect\citeauthoryear{{Ito} \& {Tanikawa}}{{Ito} \&
  {Tanikawa}}{2002}]{2002MNRAS.336..483I}
{Ito} T.,  {Tanikawa} K.,  2002, Monthly Notices of the Royal Astronomical
  Society, 336, 483

\bibitem[\protect\citeauthoryear{{Laskar}}{{Laskar}}{1989}]{1989Natur.338..237L}
{Laskar} J.,  1989, Nature, 338, 237

\bibitem[\protect\citeauthoryear{{Lehto}, {Kotiranta}, {Valtonen},
  {Hein{\"a}m{\"a}ki}, {Mikkola} \& {Chernin}}{{Lehto}
  et~al.}{2008}]{2008MNRAS.388..965L}
{Lehto} H.~J.,  {Kotiranta} S.,  {Valtonen} M.~J.,  {Hein{\"a}m{\"a}ki} P.,
  {Mikkola} S.,    {Chernin} A.~D.,  2008, Monthly Notices of the Royal
  Astronomical Society, 388, 965

\bibitem[\protect\citeauthoryear{{Leigh} \& {Wegsman}}{{Leigh} \&
  {Wegsman}}{2018}]{2018MNRAS.476..336L}
{Leigh} N. W.~C.,  {Wegsman} S.,  2018, Monthly Notices of the Royal
  Astronomical Society, 476, 336

\bibitem[\protect\citeauthoryear{{Mardling}}{{Mardling}}{2008}]{2008LNP...760...59M}
{Mardling} R.~A.,  2008, {Resonance, Chaos and Stability: The Three-Body
  Problem in Astrophysics}.
p.~59

\bibitem[\protect\citeauthoryear{{Martynova} \& {Orlov}}{{Martynova} \&
  {Orlov}}{2014}]{2014ARep...58..756M}
{Martynova} A.~I.,  {Orlov} V.~V.,  2014, Astronomy Reports, 58, 756

\bibitem[\protect\citeauthoryear{{Miller}}{{Miller}}{1964}]{1964ApJ...140..250M}
{Miller} R.~H.,  1964, Astrophysical Journal, 140, 250

\bibitem[\protect\citeauthoryear{{Orlov}, {Titov} \& {Shombina}}{{Orlov}
  et~al.}{2016}]{2016ARep...60.1083O}
{Orlov} V.~V.,  {Titov} V.~A.,    {Shombina} L.~A.,  2016, Astronomy Reports,
  60, 1083

\bibitem[\protect\citeauthoryear{{Portegies Zwart} \& {Boekholt}}{{Portegies
  Zwart} \& {Boekholt}}{2014}]{2014ApJ...785L...3P}
{Portegies Zwart} S.,  {Boekholt} T.,  2014, Astrophysical Journal Letters,
  785, L3

\bibitem[\protect\citeauthoryear{{Portegies Zwart} \& {Boekholt}}{{Portegies
  Zwart} \& {Boekholt}}{2018}]{2018CNSNS..61..160P}
{Portegies Zwart} S.~F.,  {Boekholt} T.~C.~N.,  2018, Communications in
  Nonlinear Science and Numerical Simulations, 61, 160

\bibitem[\protect\citeauthoryear{{Quinlan} \& {Tremaine}}{{Quinlan} \&
  {Tremaine}}{1992}]{1992MNRAS.259..505Q}
{Quinlan} G.~D.,  {Tremaine} S.,  1992, Monthly Notices of the Royal
  Astronomical Society, 259, 505

\bibitem[\protect\citeauthoryear{{Stone} \& {Leigh}}{{Stone} \&
  {Leigh}}{2019}]{2019arXiv190905272S}
{Stone} N.~C.,  {Leigh} N. W.~C.,  2019, arXiv e-prints, p. arXiv:1909.05272

\bibitem[\protect\citeauthoryear{{Sussman} \& {Wisdom}}{{Sussman} \&
  {Wisdom}}{1992}]{1992Sci...257...56S}
{Sussman} G.~J.,  {Wisdom} J.,  1992, Science, 257, 56

\bibitem[\protect\citeauthoryear{{Szebehely} \& {Peters}}{{Szebehely} \&
  {Peters}}{1967}]{1967AJ.....72..876S}
{Szebehely} V.,  {Peters} C.~F.,  1967, Astronomical Journal, 72, 876

\bibitem[\protect\citeauthoryear{{Tanikawa}, {Umehara} \& {Abe}}{{Tanikawa}
  et~al.}{1995}]{1995CeMDA..62..335T}
{Tanikawa} K.,  {Umehara} H.,    {Abe} H.,  1995, Celestial Mechanics and
  Dynamical Astronomy, 62, 335

\bibitem[\protect\citeauthoryear{{Urminsky}}{{Urminsky}}{2010}]{2010MNRAS.407..804U}
{Urminsky} D.~J.,  2010, Monthly Notices of the Royal Astronomical Society,
  407, 804

\bibitem[\protect\citeauthoryear{{Valluri} \& {Merritt}}{{Valluri} \&
  {Merritt}}{2000}]{2000chun.proc..229V}
{Valluri} M.,  {Merritt} D.,  2000, in {Gurzadyan} V.~G.,  {Ruffini} R.,  eds,
  The Chaotic Universe {Orbital Instability and Relaxation in Stellar Systems}.
pp 229--246

\bibitem[\protect\citeauthoryear{{Valtonen} \& {Karttunen}}{{Valtonen} \&
  {Karttunen}}{2006}]{2006Valtonen}
{Valtonen} M.,  {Karttunen} H.,  2006, {The Three-Body Problem}

\bibitem[\protect\citeauthoryear{{Wisdom}, {Peale} \& {Mignard}}{{Wisdom}
  et~al.}{1984}]{1984Icar...58..137W}
{Wisdom} J.,  {Peale} S.~J.,    {Mignard} F.,  1984, Icarus, 58, 137

\end{thebibliography}


\section*{APPENDIX A: Numerical Methods}

We adopt the Agekyan-Anosova map \citep{1967AZh....44.1261A, 1968SvA....11.1006A} and sample it uniformly with a resolution of 0.015625. This results in an ensemble of 1212 initial realizations. For a high-resolution version of the map, see the ``warrior shield'' by \citet{2008MNRAS.388..965L}.

We use the arbitrary-precision N-body code \texttt{Brutus} \citep{2015ComAC...2....2B} and vary its two main parameters, $\epsilon$, the Bulirsch-Stoer tolerance, and, $L_w$, the word-length in bits. In order to reduce the grid of parameters to vary, we fix $L_w = 1024$\,bits, which corresponds to about 300 digits, which is sufficient for the calculations in this work. 

We evolve each initial realization to the point of a permanent binary-single configuration, or a maximum time of 10,000 time units \citep{1986LNP...267..233H}. The single body is defined to be permanently escaped if 1) it is separated from the binary center of mass by more than 10 distance units, 2) it is moving away from the binary, and 3) its energy is positive. Note that for near parabolic escapes these criteria are only fulfilled at very large separations, potentially exceeding our maximum time limit. 

Once the system has fulfilled the escape criteria at time $t=T$, we reverse the velocity of each body, and continue the integration up to $t=2T$. Then, we compare the initial snapshot to the final snapshot by calculating the {Euclidean norm of the distance between the solutions in position space}. This is similar to the phase space distance defined by \citet{1964ApJ...140..250M}, but only using the position coordinates. {On the one hand, this is done to avoid confusion between adding quantities with different units, but also because from experience, we noticed that if the Euclidean norm in position space is small, then this is also the case in momentum space and vice versa. Hence, if the initial positions are retrieved after performing the reversibility experiment, then this must also be the case for the momenta. Otherwise chaos would have caused the time-reversed solution to exponentially diverge from the forward-integrated solution.} {If the Euclidean norm of the distance in position space}, $\Delta$, is sufficiently small, then the simulation has passed the reversibility test. If not, then we redo the simulation with a higher accuracy (smaller $\epsilon$), until for some accuracy the reversibility test is successful. This way, we iteratively increase the fraction of reversible solutions. Our criterion for deeming a solution reversible is: $\log_{10}\,\Delta < -3$, i.e. each position coordinate of each body in the initial and final snapshot differs only in the third decimal place or beyond. The phenomenon that, after iteratively increasing the accuracy and precision, the first $n$ decimal places of the solution have converged, and the solution has started to become time-reversible up to $n$ decimal places, is defined as {\it definitive reversibility} \citep{2018CNSNS..61..160P}.

Once we have the ensemble of definitive reversible solutions for the Agekyan-Anosova map, we measure the lifetime, $T$, and amplification factor, $A$, for each system. The lifetime is measured by considering the final snapshot of the forward integration, consisting of the permanent binary+single, and then	retracing their steps to the moment when the binary+single were closest to the moment of the final ejection. Especially for near parabolic escapes, this can cut off a significant fraction of the simulation as the escape criteria are only fulfilled when the binary and single are at very large separations. The amplification factor of a small initial perturbation is calculated by measuring the {Euclidean norm of the distance in position space} between the forward and backward integration as a function of time. The backward integration diverges exponentially from the forward integration at a rate given by the {finite-time Lyapunov exponent \citep{1991pscn.proc...47H}} of the system. Note that the perturbation is smallest at the end of the forward integration. The amplification factor is then defined as the ratio of the initial and final {Euclidean norm of the distance in position space}, $A = \Delta_T/\Delta_{0}$.

\section*{APPENDIX B: Supplementary Text}

In Fig.~1 we present our low resolution Agekyan-Anosova map, where we plot the lifetime of the triple system as a function of initial condition. We observe the blue ``rivers'' of systems that dissolve rapidly, and the surrounding chaotic landscape where nearest neighbours can have very different lifetimes. When comparing the least accurate ($\epsilon = 10^{-6}$) and the most accurate ($\epsilon = 10^{-70}$) maps, we observe that there are ``microscopic'' differences. For example the black dots, which represent very long lived systems, are differently populated in the two maps, However, in a ``macroscopic'' sense,  the maps look similar. 
To make the comparison more quantitative, we take the distributions of triple lifetime and binding energy of the final, permanent binary, and perform a two-sample Kolmogorov-Smirnoff test. We find that the least and most accurate datasets are statistically indistinguishable according to the Kolmogorov-Smirnoff test, giving p-values of 0.72 (lifetimes) and 0.85 (binding energies). These results demonstrate that, for the Agekyan-Anosova map, approximate computations are nevertheless reliable in a statistical sense \citep{1993ApJ...415..715G}, verifying the ergodic-like property of ``{\it nagh-Hoch}'' \citep{2018CNSNS..61..160P}.

\paragraph*{Correlation between amplification factor and fraction of irreversible solutions}\label{sec:result1}

In Fig.~2, we plot the fraction of irreversible solutions as a function of numerical accuracy, i.e. the Bulirsch-Stoer tolerance, $\epsilon$. We observe that initially, at low accuracy, the fraction of irreversible solutions is close to unity. As we increase the accuracy to about standard double-precision, the reversible and irreversible fractions have become roughly equal. This result is consistent with that of \citet{2008MNRAS.388..965L}. By increasing the numerical accuracy beyond machine-precision, we demonstrate that we are able to further decrease the fraction of irreversible solutions. In Fig.~2, we show that the fraction of irreversible solutions is accurately fitted by a power law, given by

\begin{equation}
\log_{10} f_{\rm{irr}} = \alpha \log_{10}\,\epsilon + \beta,
\label{eq:2}
\end{equation}  

\noindent with $\alpha=0.029 \pm 0.001$ and $\beta=0.15 \pm 0.04$. By the time we reached a Bulirsch-Stoer tolerance of $\epsilon = 10^{-70}$, the fraction of irreversible solutions had dropped to about 1 percent, which is when we decided to end the iteration for practical reasons. 

In Fig.~3 we plot the distribution of amplification factors. This distribution is also accurately fitted by a power law, given by 

\begin{equation}
\log_{10}\frac{df}{d\log_{10} A} = \gamma \log_{10} A + \delta,
\label{eq:1}
\end{equation}

\noindent with $\gamma=-0.0270 \pm 0.0008$ and $\delta=-1.20 \pm 0.01$. If this relation could be extrapolated, this would imply that for a very high sampling of the Agekyan-Anosova map, there should be some systems with amplification factors exceeding $10^{100}$, which would take a long time to calculate up to convergence. The average wall-clock time of our simulations was about 7 hours, with the longest run taking 1 month to complete.

The coefficients $\alpha$ and $\gamma$ in Eq.~3 and 4 are equal to within $3\sigma$. This suggests that there is a relation between the distribution of amplification factors and the fraction of irreversible solutions for some specified numerical accuracy. Given a Bulirsch-Stoer parameter, $\epsilon$, we are only able to resolve amplification factors of order $\epsilon^{-1}$, i.e. with an $\epsilon = 10^{-10}$ we can resolve amplification factors of order $A=10^{10}$. Hence, the fraction of irreversible solutions should approximately be equal to the fraction of systems with an $A > \epsilon^{-1}$, i.e. $f_{\rm{irr}}\left( \epsilon \right) = F\left( A > \epsilon^{-1}\right)$. Hence, using Eq.~4, we can derive the following:

\begin{equation}
f_{\rm{irr}}\left( \log_{10} \epsilon \right) = F\left( \log_{10} A > -\log_{10} \epsilon \right), \\
\end{equation}

\begin{equation}
f_{\rm{irr}}\left( \log_{10} \epsilon \right) = \int_{-\log_{10} \epsilon}^{\infty} \frac{df}{d\log_{10} A} d\log_{10} A, \\
\end{equation}

\begin{equation}
f_{\rm{irr}}\left( \log_{10} \epsilon \right) = \int_{-\log_{10} \epsilon}^{\infty} 10^{\gamma \log_{10} A + \delta} d\log_{10} A, \\
\end{equation}

\begin{equation}
f_{\rm{irr}}\left( \log_{10} \epsilon \right) \sim \left[ 10^{\gamma \log_{10} A} \right]^{\infty}_{-\log_{10} \epsilon}, \\
\end{equation}

\begin{equation}
f_{\rm{irr}}\left( \log_{10} \epsilon \right) \sim 10^{-\gamma \log_{10} \epsilon}. 
\end{equation}

\noindent Finally, taking the logarithm, we can write

\begin{equation}
\log f_{\rm{irr}} \sim -\gamma \log_{10} \epsilon.
\end{equation}

\noindent Comparing this expression to Eq.~3, we conclude that $-\gamma = \alpha$.


\label{lastpage}

\end{document}